# Unlocking the Future: A Cloud-Based Artificial Intelligence Access Control System

by Hamidreza Yaghoubi, Navtaj Randhawa (University of Applied Sciences Burgenland), and Igor Ivkić (University of Applied Sciences Burgenland and Lancaster University, UK)

*Traditional access control systems, such as key cards, PIN pads, and physical keys, face challenges in scalability, security, and user experience in today's digital world. We present a cloud-based entry system using Raspberry Pi hardware and Amazon Web Services (AWS) technologies like Lambda, Simple Storage Service (S3), and Rekognition. This solution (AWSecure Entry System) enhances security, streamlines authentication, and increases operational efficiency.*

Traditional security systems are increasingly challenged by the rapid evolution of digital threats and often rely on outdated technologies that lack the flexibility and scalability required in today's dynamic security environment. Moreover, physical keys and passwords are increasingly vulnerable to theft, counterfeiting and hacking. Furthermore, managing these systems remains labour-intensive and error-prone, leading to potential security breaches with significant financial and reputational consequences. In response to these vulnerabilities, the integration of Artificial Intelligence (AI) into security systems offers a transformative solution. AI enhances monitoring, data analysis, and threat response, especially when combined with Internet of Things (IoT) devices that continuously collect and process environmental data [1, 3]. To address the need for more flexible, scalable, and secure solutions, this article presents an Amazon Web Services (AWS)-based entry system that uses AI facial recognition technology for authentication and access control. This approach not only reduces the risk of identity theft and errors, but also ensures a faster and more secure authentication process, highlighting its potential to improve security in critical areas.

The proposed system (AWSecure Entry System) uses an IoT device (Raspberry Pi with a built-in camera and display) to capture user images and send them to the cloud. The Raspberry Pi acts as an edge device, coordinating interactions between the user interface and the AWS cloud services. To accomplish this, the Raspberry Pi uses an Application Programming Interface (API) gateway [L1] to send the captured user images to the cloud, where Lambda functions manage the flow of data and trigger the appropriate security mechanisms. One of these mechanisms is AWS Rekognition [L2], which performs real-time facial recognition by comparing the captured images with those already securely stored in S3 [2]. Another mechanism (AWS DynamoDB) [L3] provides a low-latency database for storing and retrieving user credentials. The following figure provides an overview of the proposed AWSecure Entry System architecture from the edge device (Raspberry Pi) to the cloud (AWS):

To ensure reliability and efficiency, the system shown in Figure 1 was subjected to a rigorous performance analysis, assessing facial recognition accuracy, lighting conditions, distance and angle variability. This comprehensive evaluation confirmed the robustness and adaptability of the system, demonstrating its ability to meet the stringent security requirements of modern professional environments.

The evaluation of the AWSecure Entry System was conducted through the following series of structured test scenarios to assess its performance in a variety of real-world conditions:
- Scenario 1 – Registered vs. Unregistered User Access: testing system access control for registered and unregistered individuals.
- Scenario 2 – Performance Under Different Lighting Conditions: evaluating the system's facial recognition in bright, dim and completely dark environments.
- Scenario 3 – Facial Recognition with Face Rotations: assessing system accuracy with users facing the camera at different angles (e.g. direct, 45 degrees, 90 degrees).

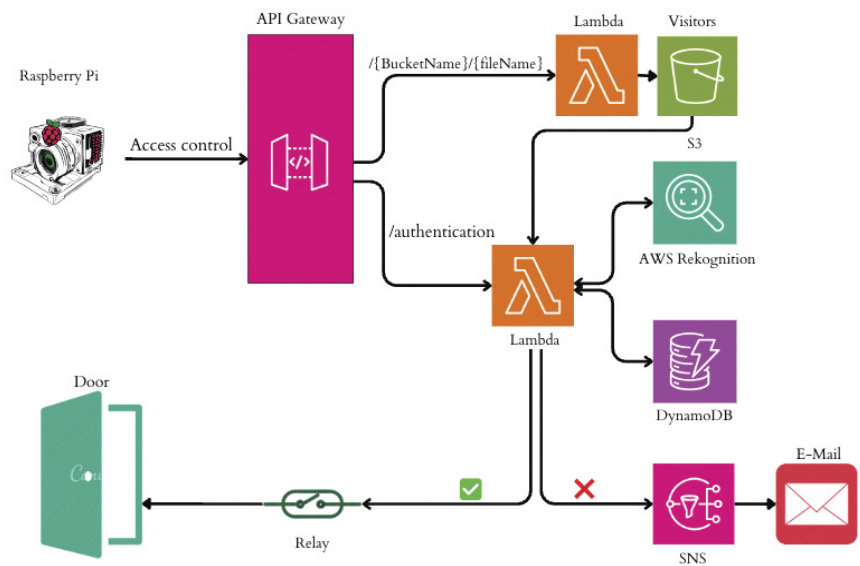

*Figure 1: AWSecure Entry System Architecture from the Edge (Raspberry Pi) to the Cloud (AWS).*

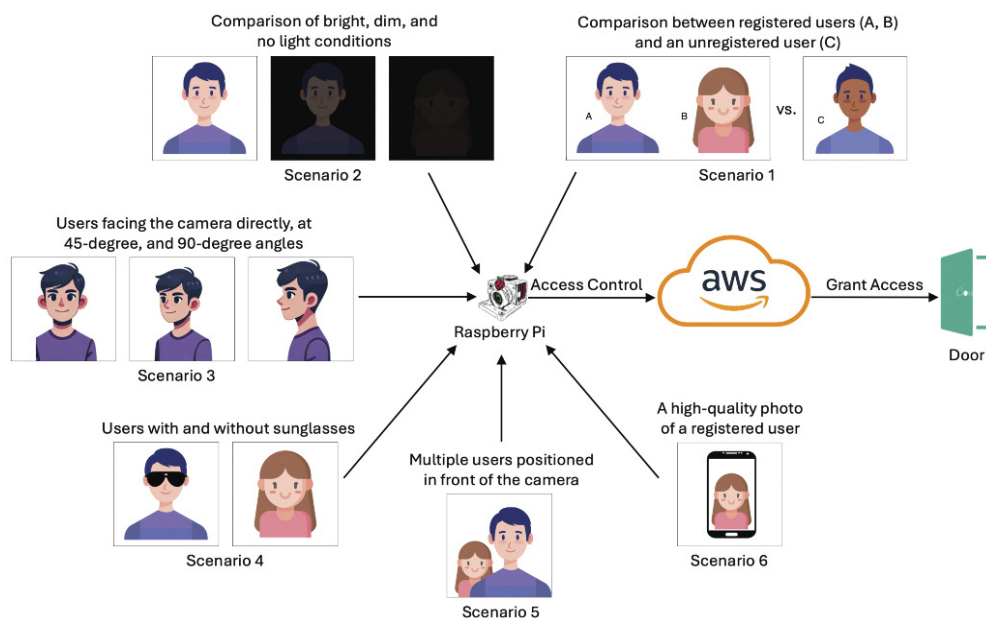

Figure 2: Evaluation of different Facial Recognition Test Scenarios using Raspberry Pi and AWS under different conditions, resulting in Door Access Granted based on user identification and matching criteria.

- Scenario 4 – Recognition with Accessories: testing the system's ability to recognise users wearing accessories such as sunglasses.
- Scenario 5 – Multi-user Recognition: evaluating how the system handles multiple users facing the camera simultaneously.
- Scenario 6 – Spoofing Test: testing the system's resilience to spoofing attempts using photos or similar deceptive techniques.

Figure 2 shows the scenarios used for the evaluation of the AWSecure Entry System.

The system successfully demonstrated its basic functionality by accurately granting access to registered users and denying access to unregistered individuals (Scenario 1). When tested under different indoor lighting conditions, the system reliably recognised users in both bright and dim environments, but failed in complete darkness, highlighting the need for adequate lighting for effective facial recognition (Scenario 2). The system also demonstrated high accuracy when users were looking directly at the camera, but struggled with significant face rotations, particularly at 45- and 90-degree angles (Scenario 3). In scenarios where users were wearing accessories (sunglasses), the system maintained its recognition ability, albeit with a slightly lower similarity score (Scenario 4). In addition, when multiple registered users were looking at the camera at the same time, the system correctly prioritised and granted access to the user closest to the camera (Scenario 5). However, a critical vulnerability was identified during the spoofing test, where the system incorrectly granted access based on a high-quality photo, indicating a susceptibility to such attacks (Scenario 6).

The results of the evaluation show that while the AWSecure Entry System performs effectively in most scenarios (Scenarios 1-2 and 4-5), particularly under controlled lighting conditions and with minor facial obstructions, it faces challenges with extreme face angles and is vulnerable to spoofing attempts (Scenarios 3 and 6). These results highlight the need for further refinement, in particular to improve the system's ability to handle different face orientations and its resistance to spoofing attacks. Despite these limitations, the PoC evaluation provides valuable insights into the system's strengths and weaknesses of the system, providing a solid foundation for future development. To ensure the robustness and reliability of the system in real-world deployments, additional layers of security, such as liveness detection and multi-angle facial recognition, should be incorporated. These enhancements would address the identified vulnerabilities and contribute to a more secure and reliable access control solution.

**Links:**
[L1] https://kwz.me/hDA
[L2] https://kwz.me/hDD
[L3] https://kwz.me/hDF

**References:**
[1] R. Anand et al., Integration of IoT with Cloud Computing for Smart Applications, CRC Press, 2023. https://books.google.at/books?id=zK3GEAAAQBAJ
[2] V. Sharma, " Object detection and recognition using Amazon Rekognition with Boto3," in 6th International Conference on Trends in Electronics andInformatics (ICOEI), pp. 727–732, 2022. https://doi.org/10.1109/ICOEI53556.2022.9776884
[3] A. Nag, J N Nikhilendra, and M. Kalmath, " IOT based door access control using face recognition," in 3rd Int. Conf. for Convergence in Technology (I2CT), pp. 1–3, 2018. https://doi.org/10.1109/I2CT.2018.8529749

**Please contact:**
Hamidreza Yaghoubi ,University of Applied Sciences Burgenland, Austria, 2310781014@fh-burgenland.at

Navtaj Randhawa, University of Applied Sciences Burgenland, Austria, 2310781020@fh-burgenland.at